\documentclass[sigconf]{acmart}
\usepackage{amsthm}
\usepackage{lineno}
\usepackage{flushend}
\usepackage{multicol}
\usepackage{threeparttable}
\usepackage{balance}
\theoremstyle{definition}
\newtheorem{definition}{Definition}[section]

\AtBeginDocument{%
  \providecommand\BibTeX{{%
    \normalfont B\kern-0.5em{\scshape i\kern-0.25em b}\kern-0.8em\TeX}}}

\usepackage{algorithm}
\usepackage{multirow}
\usepackage{float}

\copyrightyear{2022} 
\acmYear{2022} 
\setcopyright{acmlicensed}\acmConference[CIKM '22]{Proceedings of the 31st ACM International Conference on Information and Knowledge Management}{October 17--21, 2022}{Atlanta, GA, USA}
\acmBooktitle{Proceedings of the 31st ACM International Conference on Information and Knowledge Management (CIKM '22), October 17--21, 2022, Atlanta, GA, USA}
\acmPrice{15.00}
\acmDOI{10.1145/3511808.3557336}
\acmISBN{978-1-4503-9236-5/22/10}

\vspace{-0.8cm}

\title{Graph Based Long-Term And Short-Term Interest Model for Click-Through Rate Prediction}

\begin{document}




\author{Huinan Sun}
\affiliation{%
\institution{Meituan}
\city{Beijing}
\country{China}
}
\email{sunhuinan@meituan.com}

\author{Guangliang Yu}
\authornote{Huinan Sun and Guangliang Yu contributed equally to this research}
\affiliation{%
\institution{Meituan}
\city{Beijing}
\country{China}
}
\email{yuguangliang@meituan.com}

\author{Pengye Zhang}
\affiliation{%
\institution{Meituan}
\city{Beijing}
\country{China}
}
\email{zhangpengye@meituan.com}

\author{Bo Zhang}
\affiliation{%
\institution{Meituan}
\city{Beijing}
\country{China}
}
\email{zhangbo126@meituan.com}

\author{Xingxing Wang}
\affiliation{%
\institution{Meituan}
\city{Beijing}
\country{China}
}
\email{wangxingxing04@meituan.com}

\author{Dong Wang}
\affiliation{%
\institution{Meituan}
\city{Beijing}
\country{China}
}
\email{wangdong07@meituan.com}

\renewcommand{\shortauthors}{Huinan Sun et al.}




\begin{abstract}
Click-through rate (CTR) prediction aims to predict the probability that the user will click an item, which has been one of the key tasks in online recommender and advertising systems. In such systems, rich user behavior ($viz.$ long- and short-term) has been proved to be of great value in capturing user interests. Both industry and academy have paid much attention to this topic and propose different approaches to modeling with long-term and short-term user behavior data. But there are still some unresolved issues. More specially, (1) rule and truncation based methods to extract information from long-term behavior are easy to cause information loss, and (2) single feedback behavior regardless of scenario to extract information from short-term behavior lead to information confusion and noise. To fill this gap, we propose a \textbf{G}raph based \textbf{L}ong-term and \textbf{S}hort-term interest \textbf{M}odel, termed GLSM. It consists of a multi-interest graph structure for capturing long-term user behavior, a multi-scenario heterogeneous sequence model for modeling short-term information, then an adaptive fusion mechanism to fused information from long-term and short-term behaviors. Comprehensive experiments on real-world datasets, GLSM achieved SOTA score on offline metrics. At the same time, the GLSM algorithm has been deployed in our industrial application, bringing 4.9\% CTR and 4.3\% GMV lift, which is significant to the business. 
\end{abstract}

\ccsdesc[500]{Analytics and machine learning~data mining}
\ccsdesc[500]{Neural Information and knowledge processing~neural recommendation}

\keywords{Click-Through Rate Prediction; Recommendation; Long and Short-Term Interests; User Interest Modeling}



\maketitle

\section{Introduction}
Recommender systems (RS) play a key role in online services. The recommender system includes three modules: Matching, Strategy and Click-through Rate Prediction (CTR). The matching stage deploys some simple but effective recommendation algorithms (such as collaborative filtering\cite{he2018outer,he2017neural}) to pick out a small subset of relevant items from all items.The CTR stage is to predict the probability of the user clicking on the item, which is the core module of the recommender system\cite{qu2016product}. With the rapid growth of user historical behavior data, learning the intent representation of user interest through user historical behavior has been widely introduced into CTR prediction models. According to statistics, about 70\% of users' behavior sequences are longer than two hundred in our APP. Yet, the massive amount of user behavior information not only brings information gain, but also brings new problems: how to efficiently process user historical behavior sequences (including long-term behavior and short-term behavior).

The traditional processing method\cite{zhou2018deep,hidasi2018recurrent} is to truncate the user's historical behavior sequence, and only retain a certain number of user behaviors to meet the online performance. This solution is relatively simple and crude. Although it solves the performance problem of online estimation services, it brings about information loss and reduces the estimation accuracy.

In view of the insufficiency of the truncation scheme, various solutions have been proposed in the industry. One idea is to compress the user's historical behavior sequence as much as possible without losing information. The representative work is MIMN\cite{pi2019practice}. MIMN uses the UIC structure to compress and store the user's sequence of lifecycle actions. UIC embeds the user's different interest increments into a fixed-size memory  matrix that is updated with each new behavior. In this way, the computation of user modeling is decoupled from CTR prediction, avoiding the problem of online service delays. However, encoding the historical behaviors of all users into a fixed-size memory matrix results in a lot of noise in the memory cells, which cannot accurately capture user interests.

Another idea is to build a two-stage indexing scheme that filters sub-behavior sequences in real-time from a user's full historical behavior. The representative work is SIM\cite{pi2020search}, which adopts the method of category retrieval, and uses category attributes to select and locate from the user behavior sequence. Retrieving user historical behaviors based on category attributes will inevitably miss some behavioral items with different attributes but related to the current candidate, resulting in lack of user information.

In addition, users' recent behaviors carry recent interests \cite{feng2019deep}. Industrial models using different neural network architecture such as CNN, RNN\cite{zhou2019deep}, Transformer\cite{feng2019deep}, Capsule\cite{li2019multi} and attention-based \cite{zhou2018deep,feng2019deep}.  These models are often applied to certain types of action sequences, such as click sequences. However, the user's decision-making process generates various types of behaviors, such as clicks, loads, and searches. The extraction of certain types of behaviors separates the user's sequential actions and is not conducive to capturing the full intent of the user. At the same time, the behavior of users in different scenarios is not consistent. For example, the food click preferences for breakfast and lunch are completely different. The behavior information of all scenes is mixed together, and the information easily affects each other and generates noise.

In this paper, in order to maximize the use of user behavior information to improve CTR performance, we propose the GLSM algorithm, which can efficiently and accurately retrieve information from users' long-term historical behavior information. Use scenario-based modeling solutions for short-term user behavior. At the same time, a scheme of user long-term and short-term behavior fusion is proposed.

The main contributions of this paper are as follows:
\begin{itemize}
\item Long-term behavior retrieval: In view of the large amount of long-term behavior data, a new long-term behavior retrieval scheme is proposed, which utilizes graph connectivity and multi-interest center nodes to achieve efficient multi-interest soft retrieval.

\item Short-term interest extraction: Multi-scenario heterogeneous sequence modeling is used to extract users' short-term interests in different scenarios.

\item Combination of long-term and short-term interests: An interest fusion network is designed to combine long-term and short-term interests according to the user's own characteristics.
\end{itemize}
\section{Related Work}
Rich user behavior data has proven valuable in CTR prediction. The main reason is that user historical behavior (long-term and short-term) reflects user interests. If all user behavior data is added to the model, it cannot meet the performance requirements of online CTR prediction services due to the large amount. The industry has proposed some information extraction schemes (such as MIMN\cite{pi2019practice}, and SIM\cite{pi2020search}) for long-term behavior sequences. For short-term behavior sequences, models such as DIN\cite{zhou2018deep}, DIEN\cite{zhou2019deep}, BST\cite{feng2019deep}, and CapsNet\cite{li2019multi} are also proposed to extract users' short-term interests.

\subsection{Long-Term User Interest} MIMN showed that considering long-term behavior sequences in the user interest model can significantly improve the performance of the CTR model. Although long-term behavior sequences bring useful information for user interest modeling, there are also two disadvantages: it greatly increases the latency and storage burden of online service systems and the sequence contains a lot of noise. Researchers have proposed many approaches to address the challenge of modeling long-term user behavior sequences. MIMN used a fixed additional storage module NTM\cite{graves2014neural} to store the user's long-term interest vector in a compressed form, which solves the problem of a large amount of user behavior data storage. The user's long-term interest vector is updated offline in an asynchronous manner. Since there is no inference time limit offline, MIMN can theoretically model any sequence length. Yet, MIMN cannot learn various user interest vectors for different target items, resulting in information loss. SIM proposes an online two-stage retrieval method. It retrieves relevant behaviors from users' long-term behaviors based on current candidate item features such as categories. It is easy to cause information loss only through the same feature with candidate to retrieval.

\subsection{Short-Term User Interest}
Classical CTR models mainly focus on extracting user interests from short-term user behavior sequences, and various neural network architectures have been proposed, such as DIN\cite{zhou2018deep}, DIEN\cite{zhou2019deep}, MIND\cite{li2019multi}, and BST\cite{feng2019deep}. DIN emphasizes that user interests are diverse, and the user's current behavior is only related to part of the historical behavior, so an attention mechanism is introduced to capture the user's different interests in different target items. DIEN points out the temporal relationship between historical behaviors and simulates the evolution of user interests. The dynamic routing method of capsule network is introduced in MIND to learn multiple interest points of user behavior. Furthermore, inspired by self-attention in the NLP domain, Transformer is introduced in BST, which extracts deeper representations for each item and passes Transformer model using Transformer model.

To address these issues, we propose a comprehensive model for modeling users' historical behavior, called a graph retrieval-based long-term and short-term interest fusion model. In the following sections, we first introduce the GLSM algorithm framework, then briefly introduce the deployment scheme of GLSM in the industry, and finally compare GLSM with the classic CTR estimation method in the experimental part, and further do some open discussions at the end of this article.

\section{graph based long-term and short-term interest model}
The overall workflow of GLSM is shown in Figure 1. GLSM consists of three parts: a graph-based long-term interest retrieval module, a short-term multi-intent interest recognition module, and a long-term and short-term interest recognition module. In the following, we briefly introduce the CTR paradigm, and then introduce the role of each module of GLSM in CTR.

\begin{figure*}
  \centering
  \includegraphics[scale=0.47]{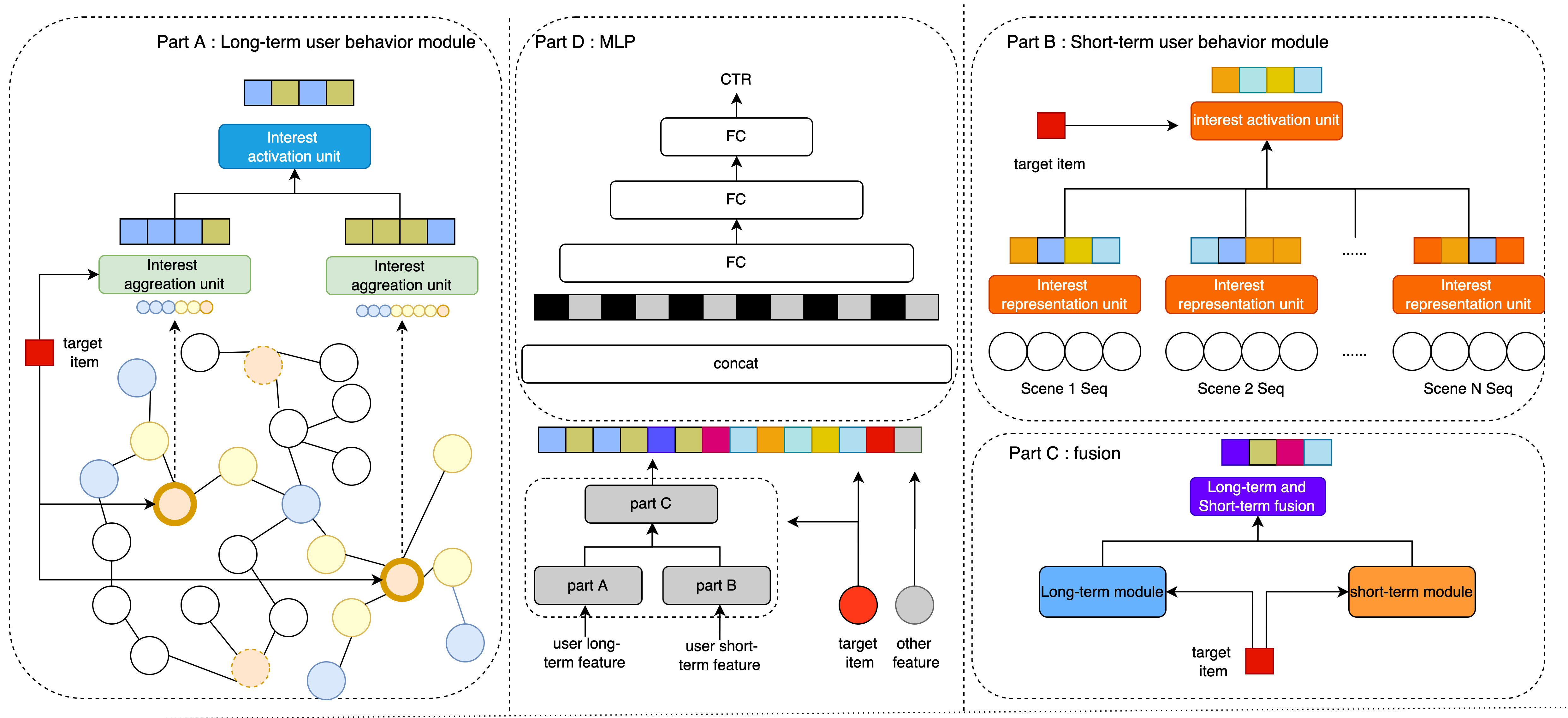}
  \caption{In the GLSM algorithm, part A is a graph-based long-term interest retrieval module; part B is the short-term multi-intent recognition interest module; part C is the long-term and short-term interest fusion module; part D is the rest model.}
  \vspace{-0.2cm}
\end{figure*}


\subsection{Click-Through Rate Prediction}
For CTR prediction task, there are M users in $U = \{ u_1, u_2, ...... , u_M \}$ and N items in $ V = \{v_1, v_2, ...... , v_N\} $. We define user behavior events as $b=(u,t,v,c,q)$. In the context of $c$, user $u$ performs $q$ action on item $v$ at time $t$. $c$ includes category, time, location, etc.

\begin{definition}[User behavior sequence]
The behavior sequence of user $u$ is $S_u = (b_{u,1},b_{u,2},b_{u,3},......,b_{u,n})$ , where $b_{u,i}$ represents the i-th user behavior of the user. 
\end{definition}

So, The CTR prediction of an user $u$ clicking on a target item $v$ is calculated via:
\begin{equation}
    y = DNN(F_e(S_u), G_e(u_p, v_p))
\end{equation}
where $F_e(S_u)$ means to extract the related behavior from $S_u$ and convert it into embedding expression, $G_e(u_p, v_p)$ represents the extraction of relevant auxiliary information from user profiles ($u_p$) and item profiles ($v_p$) with embedding representation. Generally speaking, $F_e$ is the core of the model. The reason is that $S_u$ reflect the intrinsic and multi-facet user's interests.
It is generally accepted in the industry that long-term behavior tends to reflect the user's stable interests, short-term behavior tends to include the user's early adopters and changeable intentions\cite{ren2019lifelong} \cite{pi2020search}. Therefore, this paper models long-term and short-term as two parts, so Eqn (1) is changed to:

\begin{equation}
    y = DNN(F_e^l(S_u^l), F_e^s(S_u^s), M_e(S_u^l, S_u^s), G_e(u_p, v_p))
\end{equation}

$S_u^l$ and $S_u^s$ represent the long-term and short-term parts of $S_u$, respectively, and $F_e^l$ and $F_e^s$ are the corresponding embedding expression functions. $M_e$ represents the fusion of long-term and short-term user behavior.In the following sections, we will discuss in detail the implementation of $F_e^l$, $F_e^s$, and $M_e$ in GLSM.

\subsection{Graph-based Long-Term Interest Retrieval Module}

The user's long-term behavior contains both valuable and noisy information. If it added to the model without distinction, and the irrelevant information can interfere with model estimation. Therefore, the most critical capability of the $F_e^l$ function is to efficiently and accurately retrieve relevant information from a large amount of information. Assuming that there are a total of N behaviors in the user's long-term behavior sequence, traversing whether the N behaviors are related one by one, the time complexity is O(N). This method is obviously time-consuming and labor-intensive. Therefore, there is a need to further reduce the complexity.

To tackle this challenge, we propose a graph-based retrieval structure(GRS) as the core module of $F_e^l$, as shown in Figure 2. In GRS,based on graph connectivity, efficiency, correlation, the input is the target item, and it is extended to its one-hop (or multi-hop) neighbors via the center node. These neighbors are output as the related behavior. The construction and retrieval process of the GRS will be described below.

\begin{figure}[h]
  \centering
  \includegraphics[width=\linewidth]{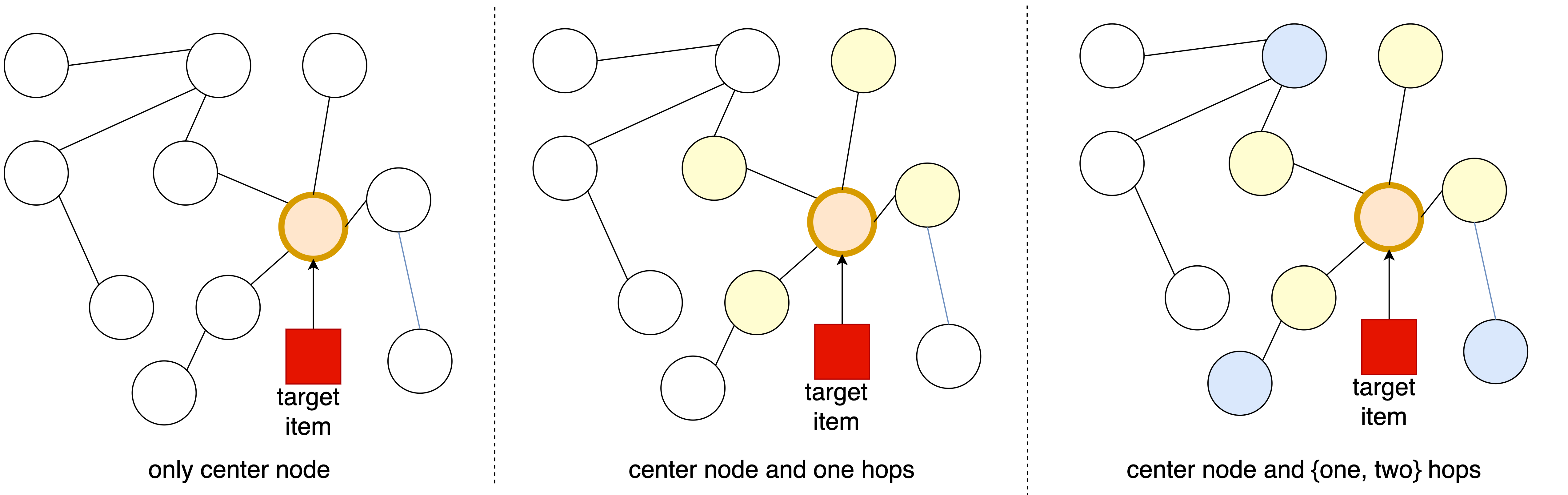}
  \caption{In GRS, the retrieval starts from the center node, and the nodes after N hops are the related nodes.}
  \vspace{-0.4cm}
\end{figure}

\subsubsection{\textbf{Graph-based retrieval structure - Construction}}

\begin{definition}(Global graph)
$G_g(global-graph) = (V, E)$. $V$ and $E$ denote node sets and edge sets. For $V$, node represents an item. For $E$, if in any $S_u$, $v_i$ and $v_j$ appear in $b_{u,i}$ and $b_{u,i+1}$, then $v_i$ and $v_j $ forms an edge. In this way, we can build a isomorphic global graph $G_g$ based on $S_u$ of all users.
\end{definition}

\begin{definition}(Local graph)
$G_l(local-graph) = (V_u, E_u)$. Compared with the global graph, A local graph is a special case of a global graph on a single user. Both $V_u$ and $E_u$ are constructed within the scope of a single $S_u$.
\end{definition}

\begin{definition}(Center Nodes)
$V_c$. In $G_l$, the center node is the key representative node of the graph.
\end{definition}

Each user has its own GRS, which consists of $G_l$ and center nodes. In GRS, the target item passes through center nodes to efficiently find its related nodes. To find center nodes from $G_l$, we measure each node's local importance($l_{im}$) and global importance($g_{im}$).

For $l_{im}$, In $G_{l}$ we measure $l_{im}$ with degree centrality\cite{zhang2017degree} as:
\begin{equation}
    l_{im} = Degree Centrality(C_{deg}(v))=\frac{d_v}{|N_{s}|-1}
\end{equation}
where $N_s$ is the set of nodes in $G_{l}$, $d_v$ is the degree of node $v$. In $G_l$, the higher the degree of the node, the more it can reflect the main interest of the user.

For $g_{im}$, $g_{im}$ mainly refers to the correlation between nodes and user interests in global. Therefore, node and user interest expression at the global level is required.To calculate  mainly refers to the correlation between nodes and user interests in global. To calculate $g_{im}$, we first construct $G_g$ using $S_u$ of all users. In $G_g$, we use a graph algorithm Graphsage \cite{hamilton2017inductive} to generate embedding expressions for each item $v$. Based on global graph embedding, K-means is used to cluster user behaviors into multiple interest clusters for each user. The reciprocal distance between each item $v$ and the k-means cluster center is taken as the global importance $g_{im}$ of the node $v$

After normalizing the global importance and local importance, the fusion result is taken as the node importance.
\begin{equation}
    union_{im} = l_{im} + g_{im}
\end{equation}
Finally, we select the top $N$ nodes $V_c=\{ v_1, v_2, ......, v_n \}$ as center nodes according to the $union_{im}$ from user graph $G_{l}$.Algorithm 1 describes this process.

\begin{algorithm}
\caption{\label{algo:lagrange} Center Nodes}
\begin{flushleft} 
1: \textbf{Input:}  all $S_u$ \\
2: \textbf{Output:} For each user, center\_node\_set ($V_c$) \\
3: Build Global Graph $G_g$ \\
4: Generate Embedding for each $v$ with GraphSage($G_g$) \\
5: \textbf{for} ($u$ in $U$): \\
6: \hspace{5mm} Build Local Graph $G_l$ \\
7: \hspace{5mm} Cluster($G_l$) to get cluster\_nucleus\_set $V_{cn\_set}$\\
8: \hspace{5mm} \textbf{for} ($v$ in $S_u$): \\
9: \hspace{11mm} $l_{im} = \frac{d_v}{| N_s|-1}$ \\
10: \hspace{1cm} $g_{im}$ = 0 \\
11: \hspace{1cm} for ($v_{cn}$ in $V_{cn\_set}$): \\
12: \hspace{15mm}  \textbf{if} ($\frac{1}{\Vert \boldsymbol{E_n} - \boldsymbol{E_{v_{cn}}} \Vert_2} > g_{im}$): \\
13: \hspace{20mm}      $g_{im} = \frac{1}{\Vert \boldsymbol{E_n} - \boldsymbol{E_{v_{cn}}} \Vert_2}$ \\
14:\hspace{1cm} $union_{im} = l_{im} + g_{im}$ \\
15:\hspace{5mm} $V_c$ = $sort_{desc}(union_{im}$)[:N] \\
16:\textbf{Return} For each user, center\_node\_set ($V_c$) \\
\end{flushleft} 
\end{algorithm}

\subsubsection{\textbf{Graph-based retrieval structure - Retrieval}}
For center nodes $V_c$, they are part of the nodes in the user graph $G_l$. So, We find the top $K$ most relevant nodes to the target item from the center nodes, then we can extend the top $K$ center nodes to the neighbor nodes $V_{next\_hop\_nodes}=\{ v_{c,1}, v_{c,2}, v_{c,3} ...... \} $ according to the connectivity of the graph. Further, We can expand each element of $V_{next\_hop\_nodes}$ in the same way to get second-hop neighbors. We can expand as many hops as we want, in this study, we get second-hop neighbors of the center nodes. Algorithm 2 describes this process.

\begin{algorithm}
\caption{\label{algo:lagrange} Retrieval In Graph}
\begin{flushleft} 
1: \textbf{Input:}  center\_node\_set ($V_c$), center\_node\_topk ($K$), \\ 
\hspace{12mm} retrieval\_hop ($L$), target\_item ($v_t$)\\
\hspace{12mm} find neighborhood function $N$\\
2: \textbf{Output:} all hop neighbor nodes ($V_{nodes}$) \\
3: \textbf {for}($v_c$ in $V_c$): \\
4: \hspace{5mm} distance = $\Vert \boldsymbol{E_{v_c}} - \boldsymbol{E_{v_{t}}} \Vert_2 $ \\
5: \hspace{5mm} V\_Dis <- ($v_c$, distance) \\
6: $V_{c_k}$ = $sort\_by\_distance_{desc}$(V\_Dis)[:$K$] \\
7: \textbf {for} ($v_c$ in $V_{c_k}$): \\
8: \hspace{5mm} \textbf{for} ($l$ in $L$): \\
9: \hspace{1cm} $V_{nodes}$  <-  $N$($v_c^l$) \\
10:\textbf{Return $V_{nodes}$} \\
\end{flushleft} 
\end{algorithm}

\subsubsection{\textbf{Long-term user behavior interest aggregation unit}}

\begin{figure}[h]
  \centering
  \includegraphics[width=\linewidth]{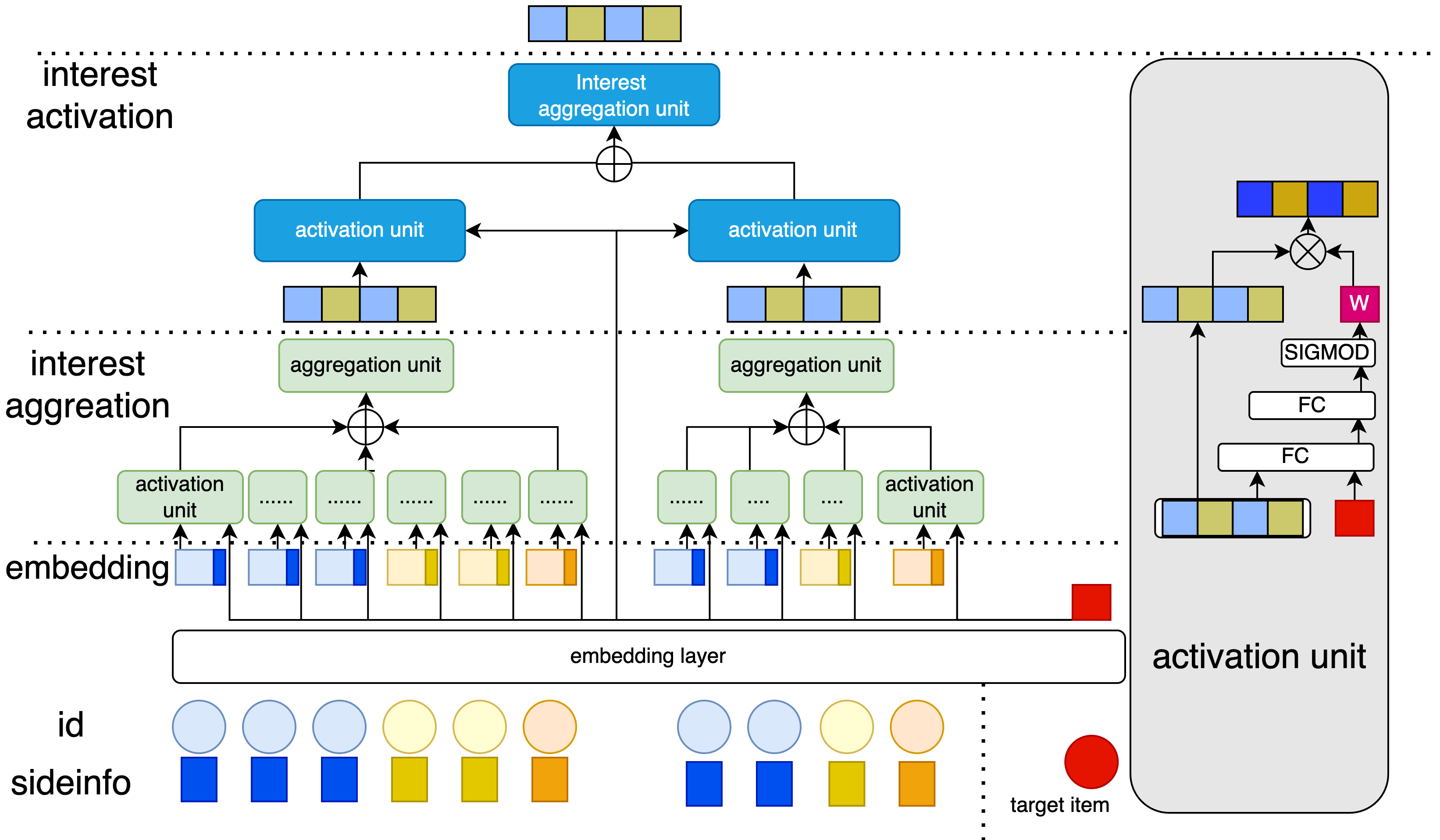}
  \caption{Long-term user behavior modeling}
  \vspace{-0.3cm}
\end{figure}

The center node $V_c$ represents the user interest cluster, and its neighbor nodes are the precise description of the user cluster. So, we aggregate all the neighbor nodes of the center node together through the attention mechanism.

Specifically, in the aggregation process, in addition to the Embedding of the neighbor node itself, the influence of the behavior sideinfo of this node is also considered: 
\begin{equation}
\begin{aligned}
&\boldsymbol{E}_{v_i} = \boldsymbol{E}_{v_i} + \boldsymbol{E}_{v_{i, sideinfo}} \\
&\boldsymbol{E}_{v_{i, sideinfo}} = \boldsymbol{E}_{{cate}_i} + \boldsymbol{E}_{{behaviortype}_i} + \boldsymbol{E}_{{discrete\_time}_i}
\end{aligned}
\end{equation}

sideinfo includes categories, behavior types, and time. Categories and behavior types map IDs to embeddings by way of embedding lookup. Due to the continuity of temporal features, we discretize time for looking up embedding:

\begin{equation}
    \begin{aligned}
        \boldsymbol{E}_{{discrete\_time}_i} = E(int(log(t_{now} - t_{v})))
    \end{aligned}
\end{equation}
where $t_{now}$ represents the current time, $t_{v}$ represents the behavior time of the current node, and the time difference represents the influence of the node over time.

The influence of $E_v$ on the target item is different. For example, when predicting the user's preference for the barbecue restaurant (target item), the relevant information of barbecue, beer, etc. is more important. Therefore, we calculate the influence weight of $E_v$:
\begin{equation}
    \begin{aligned}
\alpha_i &= attention(\boldsymbol{E}_{v_i}, \boldsymbol{E}_{v_t}) \\
&= \sigma (\boldsymbol{W}_1(\boldsymbol{W}_2([\boldsymbol{E}_{v_i}, \boldsymbol{E}_{v_t}])))
    \end{aligned}
\end{equation}

where the $\sigma$ is the sigmoid activation function, $\boldsymbol{E}_{v_t}$ is the embedding of target item and $[A, B]$ means concat two vectors. By aggregating the neighbor nodes of the center node, the representation of the current center node can be obtained. Because we have done personalized weights on neighbor nodes, we use pooling aggregation here.

\begin{equation}
    \begin{aligned}
\boldsymbol{E}_{center} = \sum_{i=0}^n \alpha_i * \boldsymbol{E}_{v_i} \\
    \end{aligned}
\end{equation}

\subsubsection{\textbf{Long-term user behavior interest activation unit}}

We have calculated the basic representation of the interest cluster. To distinguish the influence of different interest clusters on the target item, we use an attention mechanism to calculate the contribution of each interest cluster to CTR prediction. Similar as Eqn(7) and Eqn(8),We calculate the weight $\beta_j$ for each center node:
\begin{equation}
    \begin{aligned}
    \beta_j &= attention(\boldsymbol{E}_{center_i}, \boldsymbol{E}_{v_t}) \\
                    &= \sigma (\boldsymbol{W}_1(\boldsymbol{W}_2([\boldsymbol{E}_{center_i}, \boldsymbol{E}_{v_t}])))
    \end{aligned}
\end{equation}
Then we aggregate multiple center nodes by weight $\beta_j$:
\begin{equation}
    \begin{aligned}
    \boldsymbol{E}_{long} = \sum_{j = 0}^k \beta_j * \boldsymbol{E}_{center_{j}}
    \end{aligned}
\end{equation}
The complete model diagram is shown in the figure 3.

\subsection{Short-term multi-intent recognition module}

User decision-making is a continuous process in the scene \cite{chen2019reinforcement}. Decision-making process include click, cart, favorite, order, etc. Extracting only specific types of subsequences such as $S_{u}(click)$ or $S_{u}(cart)$ will lead to the splitting of scene decision-making process, which is not conducive to capturing the user's full intent. Furthermore, if the model mixes all scene behavior information, the information in different scenes interacts with each other and brings noise to the model. So we model short-term behavior as shown in Figure 4.

\begin{figure}[h]
  \centering
  \includegraphics[scale=0.47]{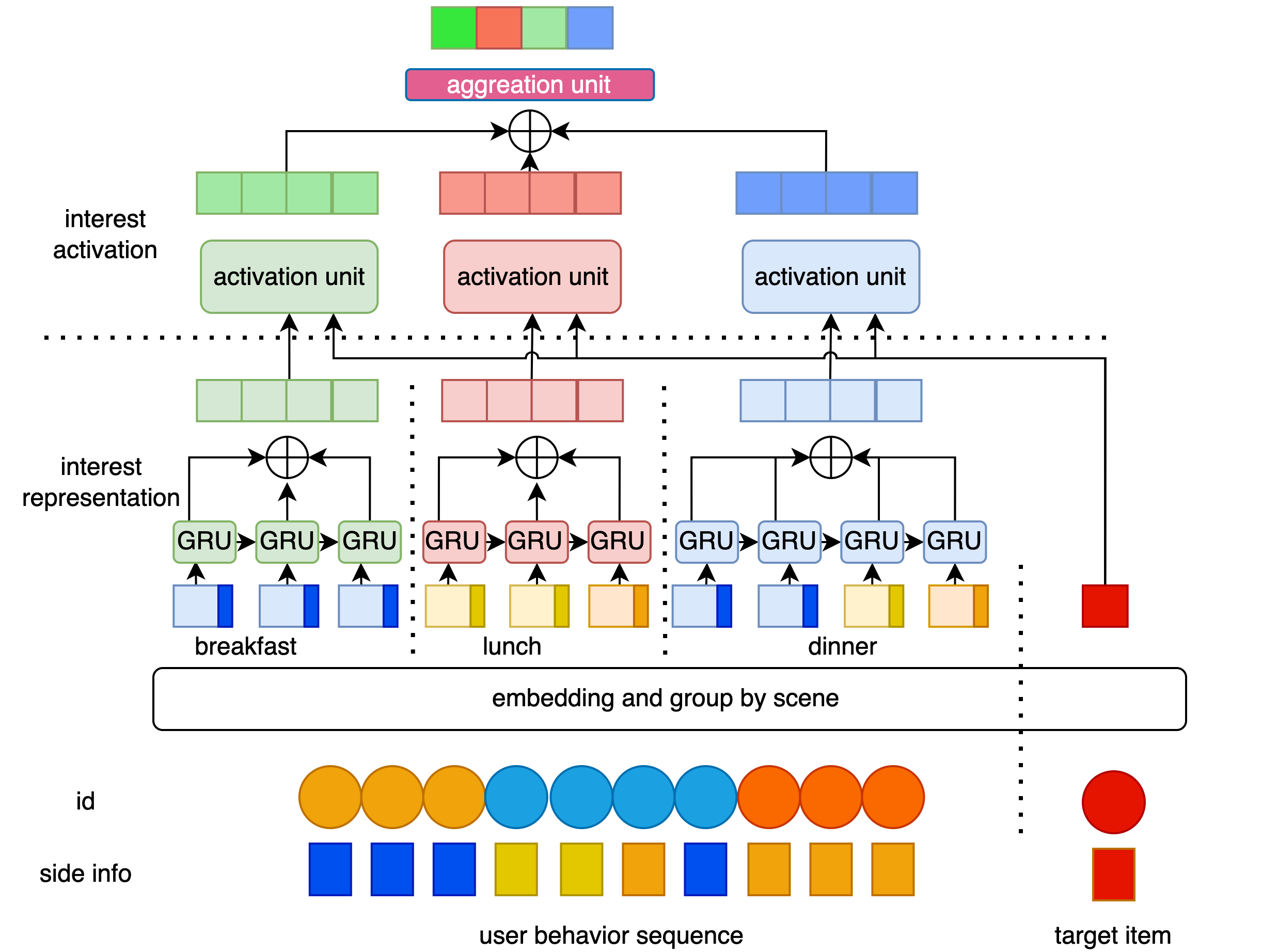}
  \caption{Short-term multi-intent recognition module}
  \vspace{-0.4cm}
\end{figure}

\subsubsection{\textbf{Short-term interest representation unit}}
Scenarios are the key to short-term scenario-based models. Generally speaking, time and location are good criteria for scene division.For example, in catering platform, user behavior is greatly affected by meal segments, behaviors at different meal segments are quite different. We divide user short-term behaviors $S_u^s $ are into multiple segments:
\begin{equation*}
    \begin{aligned}
        &S_u^s(Breakfast) = (b_{breakfast\_click},b_{breakfast\_order})\\  
        &S_u^s(Lunch) = (b_{lunch\_click},b_{lunch\_add\_cart},b_{lunch\_click})\\
        &S_u^s(Supper) = (b_{supper\_click},b_{supper\_add\_cart},b_{supper\_click})\\
    \end{aligned}
\end{equation*}
Since we want to capture behavioral temporal changes while taking into account model performance, we model users' short-term interests using continuous behaviors within scenario via an attention-based GRU model.


\begin{equation}
    \begin{aligned}
    &\boldsymbol{E}_{v_i} = \boldsymbol{E}_{v_i} + \boldsymbol{E}_{v_{i,sideinfo}} \\
    &z_t = \sigma (\boldsymbol{W}_z[\widetilde{\boldsymbol{E}}_{v_{i - 1}}, \boldsymbol{E}_{v_i}] \\
    &r_t = \sigma (W_r[\widetilde{\boldsymbol{E}}_{v_{i - 1}}, \boldsymbol{E}_{v_i}] \\
    &h_t = tanh(W[r_t \odot \widetilde{\boldsymbol{E}}_{v_{i - 1}}, \boldsymbol{E}_{v_i}] \\
    &\widetilde{\boldsymbol{E}}_{v_i} = (1 - z_t) \odot \widetilde{\boldsymbol{E}}_{v_{i - 1}} + z_t \odot h_t
    \end{aligned}
\end{equation}

where $\odot$ is element-wise multiplication.The user's decision-making behavior is continuous in the scene. If we only use one GRU in scene sequence, we can only get the final state of the user and cannot capture the intermediate state. Therefore, we retain the state of each action, and aggregate these states as the scene representation.

\begin{equation}
    \begin{aligned}
        \boldsymbol{E}_{scene} = \sum_i^n\widetilde{\boldsymbol{E}}_{v_i}
    \end{aligned}
\end{equation}

\subsubsection{\textbf{Short-term interest activation unit}}
The user's interest distribution in different scenarios has different effects on the current target item. For example, interest in breakfast (soy milk) has little effect on the user's desire to eat barbecue at night. So, according to the current scene, candidate item is used as target to activate each short-term intention $E_{scene}$ by the attention mechanism. Thus, the short-term intentions of users in different scenarios will be introduced into the model with different importance:

\begin{equation}
    \begin{aligned}
        \gamma_i &= attention(\boldsymbol{E}_{scene_i}, \boldsymbol{E}_{v_t}) \\
                    &= \sigma (\boldsymbol{W}_1(\boldsymbol{W}_2([\boldsymbol{E}_{scene_i}, \boldsymbol{E}_{v_t}])))
    \end{aligned}
\end{equation}

After calculating the Embedding expression of each scene, aggregate the Embedding of the scene to generate a short-term Embedding expression:

\begin{equation}
    \begin{aligned}
    \boldsymbol{E}_{short} = \sum_i^m \gamma_i * \boldsymbol{E}_{scene_i}
    \end{aligned}
\end{equation}

\subsection{Long-Term and Short-Term Interests Fusion Module}
The effect of long-term and short-term behavior on user decisions is individualized by the user. The long-term behavior of users with more stable interests has a greater impact on the representation of user interests, while the short-term behavior of users who like early adopters can better reflect user interests. So, we propose a user-personalized network to fuse users' long-term and short-term interests as shown in Figure 5.

\begin{figure}[h]
  \centering
  \includegraphics[scale=0.31]{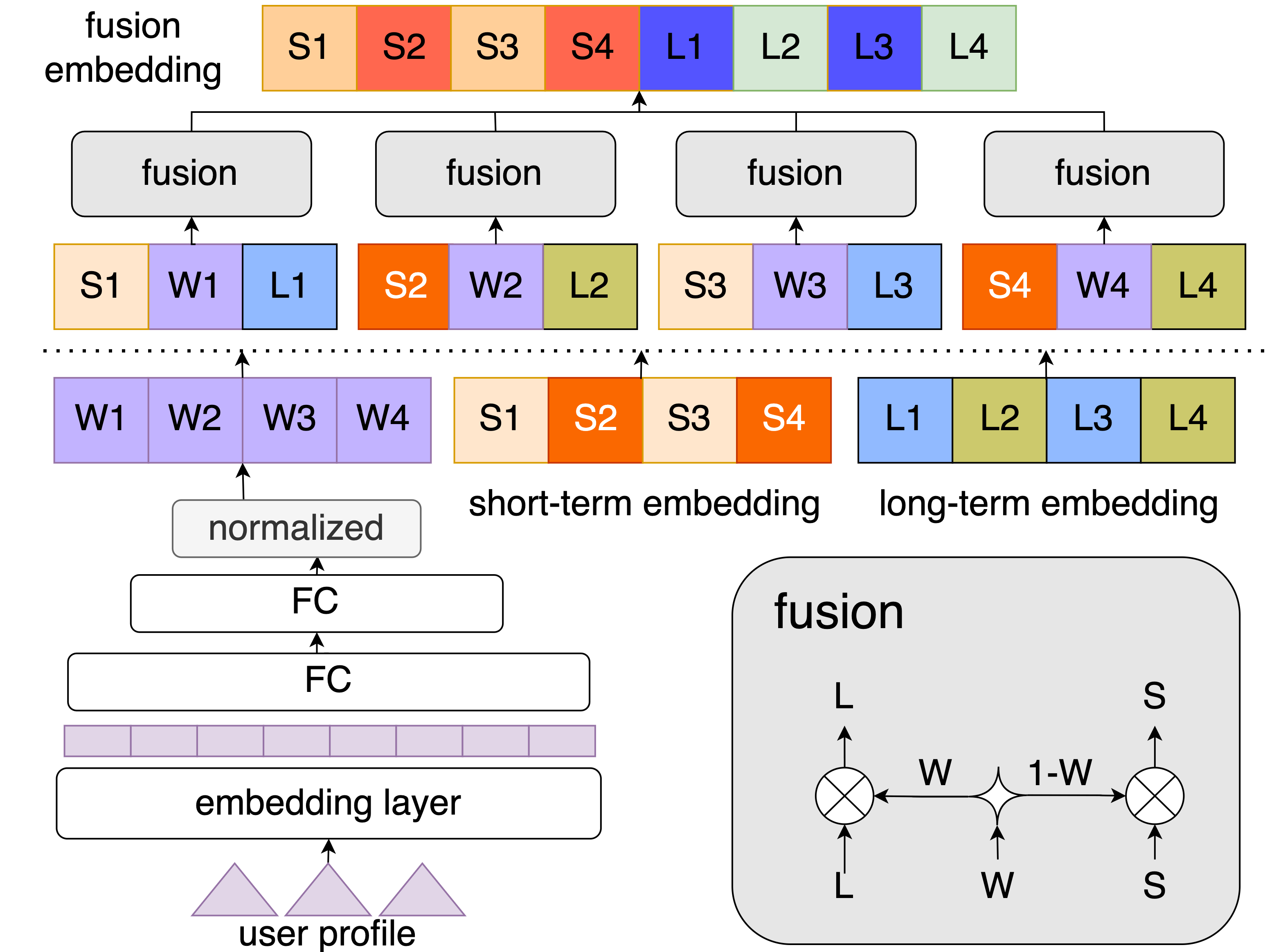}
  \caption{Long-term and short-term fusion module}
  \vspace{-0.3cm}
\end{figure}

\subsubsection{\textbf{User personalized gate network}}
User profiles is the input to the gated neural network $Gate_u$. The output is gated Embedding through multi-layer $E_{gate}$: 
\begin{equation}
    \begin{aligned}
    \boldsymbol{E}_{gate_{u}} = \boldsymbol{W}_1 (\sigma(\boldsymbol{W}_2 (\boldsymbol{E}_{\text{user profile}})))
    \end{aligned}
\end{equation}

To balance the weight influence of each interest component, $\boldsymbol{E}_{gate_u}$ is normalized, for each dimension of $\boldsymbol{E}_{gate_u}$:

\begin{equation}
    \begin{aligned}
        \widetilde{\boldsymbol{E}}_{gate_{u}}^i = \frac{exp(\boldsymbol{E}_{gate_{u}}^i)}{\sum_j^d exp(\boldsymbol{E}_{gate_{u}}^j)}
    \end{aligned}
\end{equation}

$\boldsymbol{E}_{gate_{u}}$'s embedding dimension is equal to long-term interest embedding and short-term interest embedding. It should be noted that the embedding of all input user profiles of the network does not accept the back-propagation gradient of $Gate_u$. The purpose of this operation is to reduce the impact of $Gate_u$ on the convergence of existing feature embeddings.  By adding a personalized bias term to the input of the neural network layer through $Gate_u$, computed by 

\begin{equation}
    \boldsymbol{E}_u = concat(\widetilde{\boldsymbol{E}}_{gate_{u}}\odot \boldsymbol{E}_{long}, \hspace{2mm} (1-\widetilde{\boldsymbol{E}}_{gate_{u}})\odot \boldsymbol{E}_{short})
\end{equation} 
where $\odot$ is element-wise multiplication. With the personalized $E_u$, the target prediction ability of the model can be improved.

\section{IMPLEMENTATION FOR ONLINE SERVING}
In this section, we will introduce our practical experience of implementing GLSM in our industrial application.

\subsection{Online Performance Challenges}
Industrial recommender systems need to process each traffic request in tens of milliseconds. Requests need to be processed through key processes such as match and CTR, as well as other business rules including material serving and category filtering, etc. At the same time, the amount of user's historical behavior data is large, further increasing the load of the system. The CTR module needs to efficiently and quickly get important information from a large amount of historical operational information within a limited time. This is the main challenge in deploying online systems.

\subsection{Implementation Scheme of Online Service System Based on Graph Retrieval}
\begin{figure}[h]
  \centering
  \includegraphics[scale=0.28]{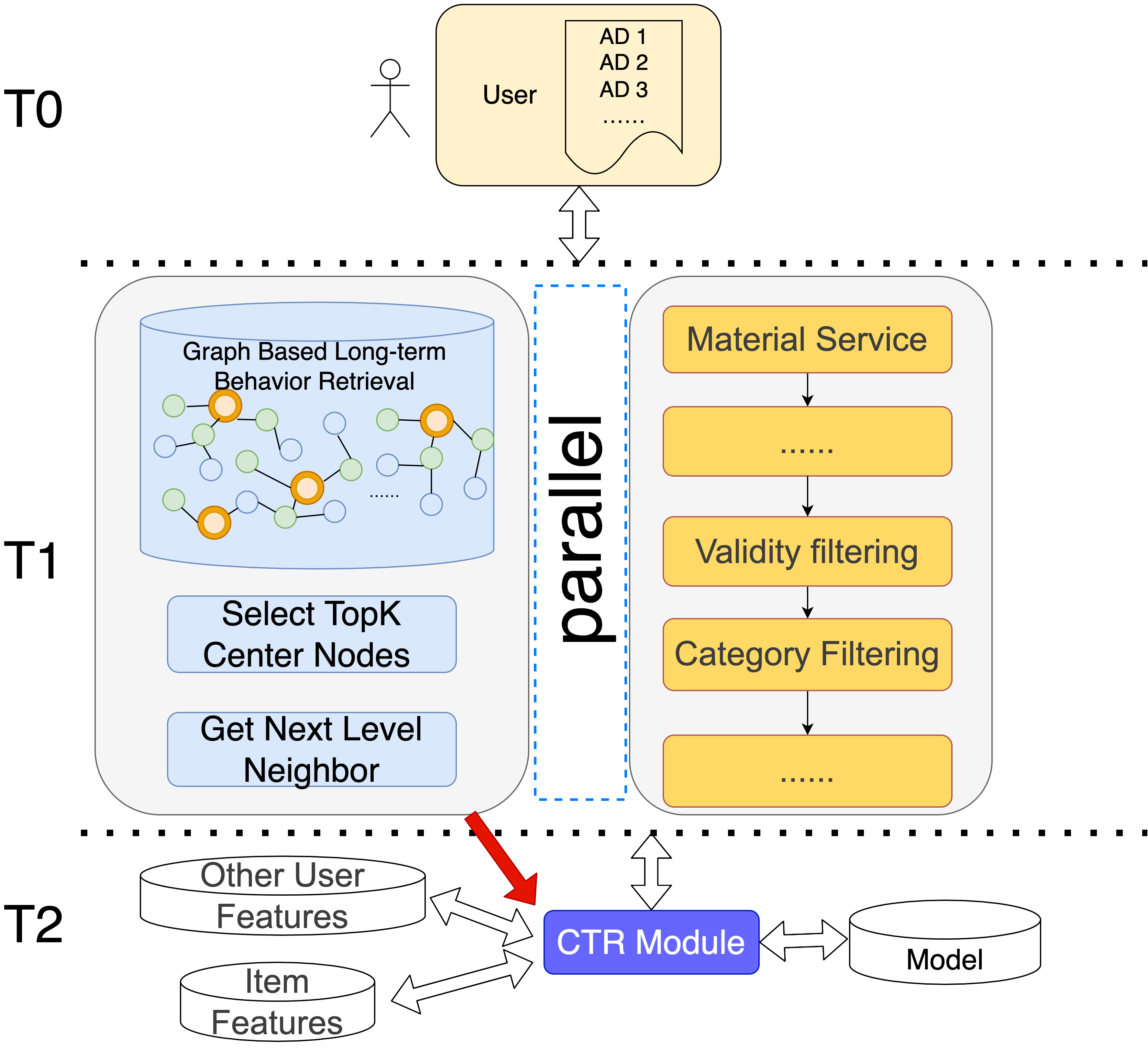}
  \caption{CTR prediction system with proposed GLSM model.The graph based long-term behavior retrieval will run in parallel with material service, saving most of latency cost for online serving.}
  \Description{}
  \vspace{-0.5cm}
\end{figure}

\textbf{Offline storage part:} To reduce the load pressure of online feature acquisition, we pre-convert users' long-term behaviors into a subgraph structure, which centered on the center nodes, and store them offline. In this way, during online retrieval, nodes retrieval can be performed directly, which saves time for construction.

\textbf{Architecture optimization:}As shown in Figure 6, in order to further improve the retrieval efficiency of users' long-term behavior, we prepend the retrieval process before the regular CTR module, and run in parallel with some other intermediate processes (material service, category filtering, etc.) In CTR module, only some other general features (user portraits, etc.) need to be obtained here.

\section{EXPERIMENTS}
In this section, we conduct experiments with the aim of answering the following three research questions:
\begin{itemize}
\item RQ1: Does our GLSM model outperform the baseline model?
\item RQ2: How does each part of our GLSM model work?
\item RQ3: What is the impact of the different components in GLSM?
\end{itemize}

Before presenting the evaluation results, we first introduce the dataset, baseline model, metrics, and experimental setup.

\subsection{DataSet}
We adopt public datasets and industrial datasets to comprehensively compare GLSM models and baseline models. The statistics of the datasets are shown in Table 1.

\begin{table}[h]
  \caption{\textcolor{black}{Statistics of datasets used in this paper}}
  \label{tab:freq}
  \centering
  \setlength{\abovecaptionskip}{0.cm}
  \setlength{\belowcaptionskip}{-0.cm}
  \begin{tabular}{c|c|c|c}
    \toprule
    Dataset & Users & Items & Instances \\
    \midrule 
     Taobao Dataset & 987994 & 4162024 & 1481728 \\
     Industrial Dataset & 102803105 & 12339496 & 415013076 \\
  \bottomrule
\end{tabular}
\end{table}


\textbf{Taobao dataset}: This dataset was first released by Alibaba-Taobao and is widely used as a common benchmark in CTR estimation tasks. It is a user behavior log for Taobao mobile application, including click, purchase, add-to-cart and favorite behaviors. There are approximately 101 actions per user, and an average of 24 actions per item, bringing the total number of actions to 100 million. We choose the closest 30 behavior as the short-term user behavior sequence, the others as long-term behavior.

\textbf{Industrial dataset}: This dataset is an industrial dataset collected by our own App which is one of the top-tier mobile Apps in our country. It is much larger than the Taobao public data set, and the maximum user behavior sequence can reach 5000. In our business, taking into account the characteristics of the business, we use the behavior of users in the past two weeks as short-term behavior, no more than 80, and the rest are long-term behavior. 

\begin{table*}[t]
  \caption{Comparisons of different models on Taobao and Industrial datasets}
  \label{tab:freq}
  \centering
  \setlength{\abovecaptionskip}{0.cm}
  \setlength{\belowcaptionskip}{-0.cm}
  \begin{threeparttable}
  \begin{tabular}{c|c|c|c|c|c|l}
    \toprule
    \multirow{2}*{Methods} & \multicolumn{3}{c}{Taobao Dataset} & \multicolumn{3}{c}{Industrial Dataset}\\
    & AUC & GAUC & Logloss & AUC & GAUC & Logloss\\
    \midrule 
     DIN + short & 0.9632 & 0.9620 & 0.1095 & 0.6543 & 0.5641 & 0.5354\\
     DIEN + short & 0.9653 & 0.9635 & 0.1079 & 0.6551 & 0.5647 & 0.5337\\
     GLSM + short & 0.9663\tnote{*} & 0.9640\tnote{*}  & 0.1055\tnote{*} & 0.6563 & 0.5657 & 0.5312\\
     DIN + long & 0.9599 & 0.9589 & 0.1149 & 0.6578 & 0.5669 & 0.5284 \\
     DIEN + long & 0.9619 & 0.9605 & 0.1143 & 0.6565 & 0.5658 & 0.5310 \\
     SIM + long & 0.9716 & 0.9694 & 0.0950 & 0.6623 & 0.5715 & 0.5234 \\
     GLSM + long & 0.9819 & 0.9811 & 0.0785 & 0.6671 & 0.5758 & 0.5163 \\
     GLSM+short+long & 0.9830\tnote{*} & 0.9820\tnote{*} & 0.0771\tnote{*} & 0.6705 & 0.5786 & 0.5178 \\
  \bottomrule
\end{tabular}
 \begin{tablenotes}
    \footnotesize
    \item[*] Due to the lack of scene-related descriptions in public datasets, we model the intent of short-term behaviors under a single scenario in public datasets  
\end{tablenotes}
\end{threeparttable}
\end{table*}

\subsection{Baselines and Metrics}

\subsubsection{Baseline}: We evaluate the performance of GLSM against the following state-of-the-art CTR methods.

\begin{itemize}
    \item DIN\cite{zhou2018deep}: Based on the attention mechanism, the user behavior sequence is given different weights through the relevance of the behavior to the target advertisement. 


    \item DIEN\cite{zhou2019deep}: Based on DIN, an interest extraction layer is designed to obtain timely interests from user behavior sequence. Meanwhile, an interest evolution layer is proposed, which uses GRU with attention update gates to simulate the interest evolution process related to the target item.

    \item  SIM\cite{pi2020search}: SIM proposes a two-stage retrieval model for long-term user behavior. It retrieves the user's long-term behavior according to the category id, selects the top K historical behaviors of the same category as the candidate, and then adds it to the model through attention mechanism. In this study, we choose the model served online in SIM paper as our baseline.

\end{itemize}

\subsubsection{Metrics}: We evaluate the CTR prediction performance with three widely used metrics. The first one is area under ROC curve (AUC) which reflects the pairwise ranking performance between click and non-click samples. In addition, in the recommender system the comparison of item rankings within the same user is a more important indicator. so we introduce GAUC\cite{zhu2017optimized} to observe the model's ability to rank different items for the same user. The other metric is log loss. Log loss is to measure the overall likelihood of the test data and has been widely used for the classification tasks. At the same time, We deploy models online in real industrial systems and use CTR as an evaluation metric in our A/B tests.

\subsection*{RQ1: Does our GLSM model outperform the baseline model?}

We evaluate the performance of GLSM and baseline models on two datasets. From Table 2, we have the following observations:

\textbf{Long-term behavior analysis}. GLSM+long outperforms the other two models because there are a large number of actions in long-term action sequences, which are not related to the target item. These behaviors bring noise to the model. DIN/DIEN+long does not specifically deal with this part of the noise information, So its performance is lowest among the three models. SIM+long achieves the purpose of filtering noise by only retrieving items of the same category as the target item. But its filter condition is a hard match, which leads to excessive noise removal. For example, barbecue and beer are common pairing categories, and hard filtering of categories will lose the matching between categories. Therefore, a complete retrieval scheme should filter noise from the long-term behavior of a large number of users and find effective information. It can be seen from the results that the effect of GLSM+long is better than that of the model SIM + long, indicating that the long-term behavior based on graph retrieval can remove redundant noise and enhance the information gain to improve the prediction effect.

\textbf{Short-term multi-intent recognition}. We compared DIN+short, DIEN+short and GLSM+short, the effect is DIN < DIEN < GLSM. DIEN captures user interest transfer through time series models such as GRU, which improves the model performance, but does not solve the phenomenon of multi-action types and multi-intent mixing in short-term behavior sequences, while GLSM not only identifies user interest transfer through multi-intent GRU modules, but also make the user's intention clearer and split out to further improve the effect. In our business, better modeling of short-term behavior can effectively capture users' recent interests and improve user experience.

\begin{figure}[h]
  \vspace{-3mm}
  \begin{minipage}[t]{0.5\linewidth}
    \centering
    \includegraphics[scale=0.22]{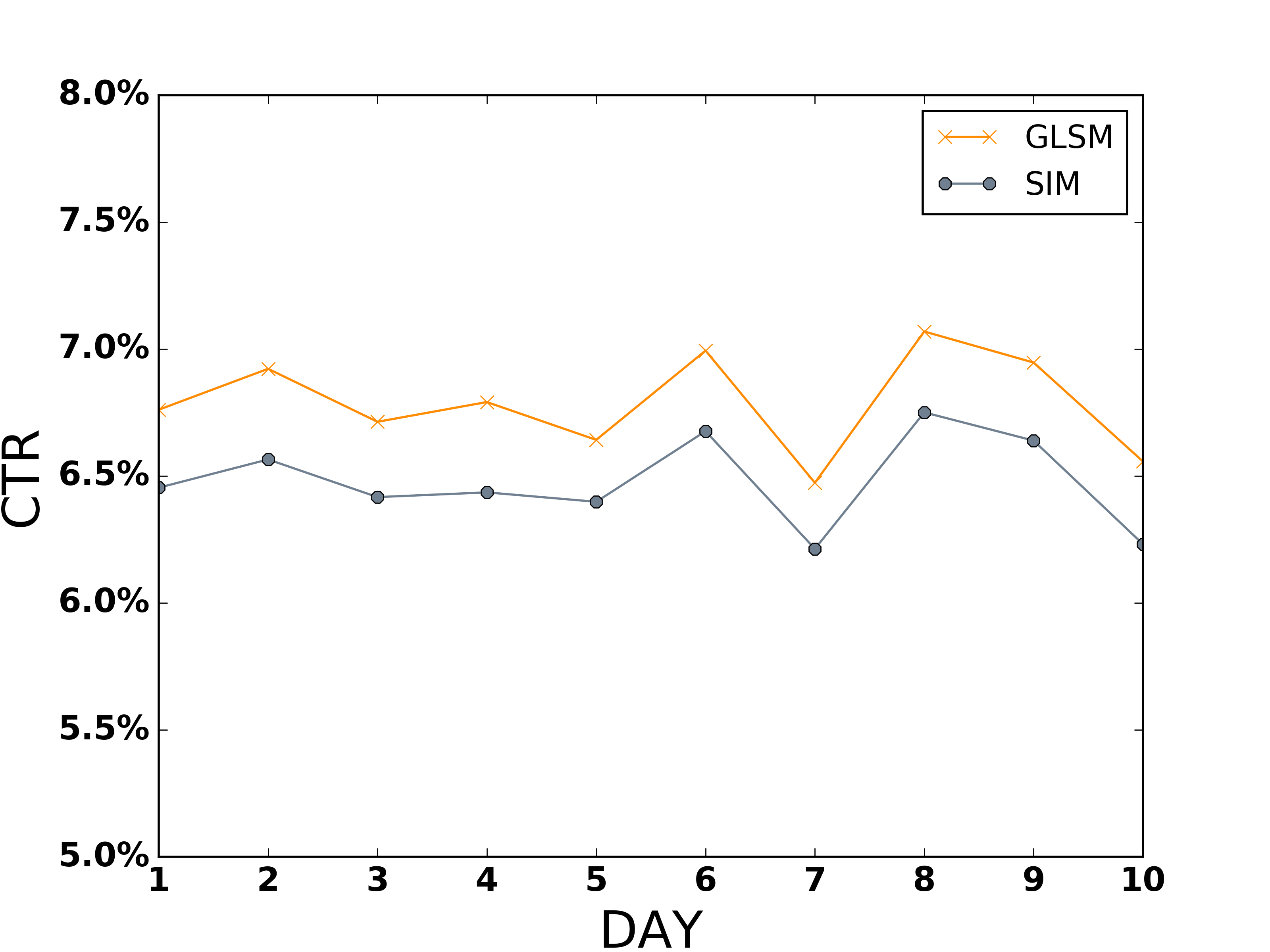}
    \label{fig:side:a}
  \end{minipage}%
  \begin{minipage}[t]{0.5\linewidth}
    \centering
    \includegraphics[scale=0.22]{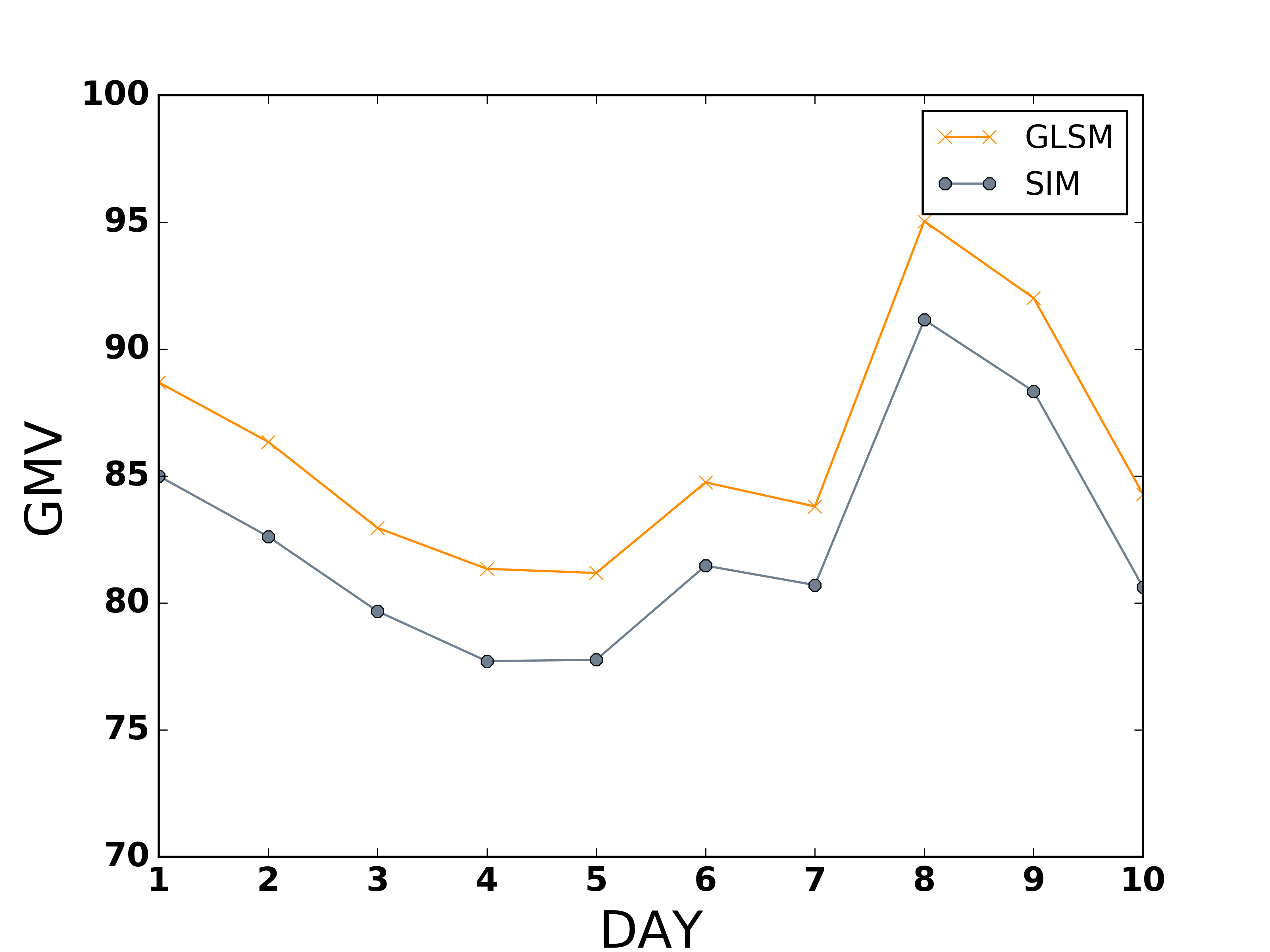}
    \label{fig:side:b}
  \end{minipage}
  \caption{Comparison of SIM and GLSM online.}
\end{figure}

\textbf{Online performance}. We deploy SIM and GLSM models on a real industrial recommender system, From figure 7 we can see that the most important online metrics CTR ($CTR=\frac{\#click}{\#pages}$) and GMV ($GMV=1000 * \frac{\#pay \, amount}{\#pages}$) have an overall average improvements of 4.9\%, 4.3\% respectively, which are significant improvement on our platform.

\subsection*{RQ2: How does each part of our GLSM model work?}

\subsubsection*{\textbf{A. Graph-based long-term interest retrieval module}}
\ 

\indent \textbf{Number of clusters}: When calculating $g_{im}$, it involves using clustering to capture the long-term interest clusters of users. Different numbers of long-term interest clusters will affect the $g_{im}$ weight calculation of each node, which in turn affects the selection of the final center node. So, we tried multiple numbers of clusters on user long-term behavior sequences, and use the Silhouette Coefficient: 
\begin{equation}
   s=\frac{b-a}{max(a,b)} 
\end{equation}
(a: The mean distance between a sample and all other points in the same class, b: The mean distance between a sample and all other points in the next nearest cluster) to measure the clustering effect.

\begin{figure}[h]
  \centering
  \includegraphics[scale=0.3]{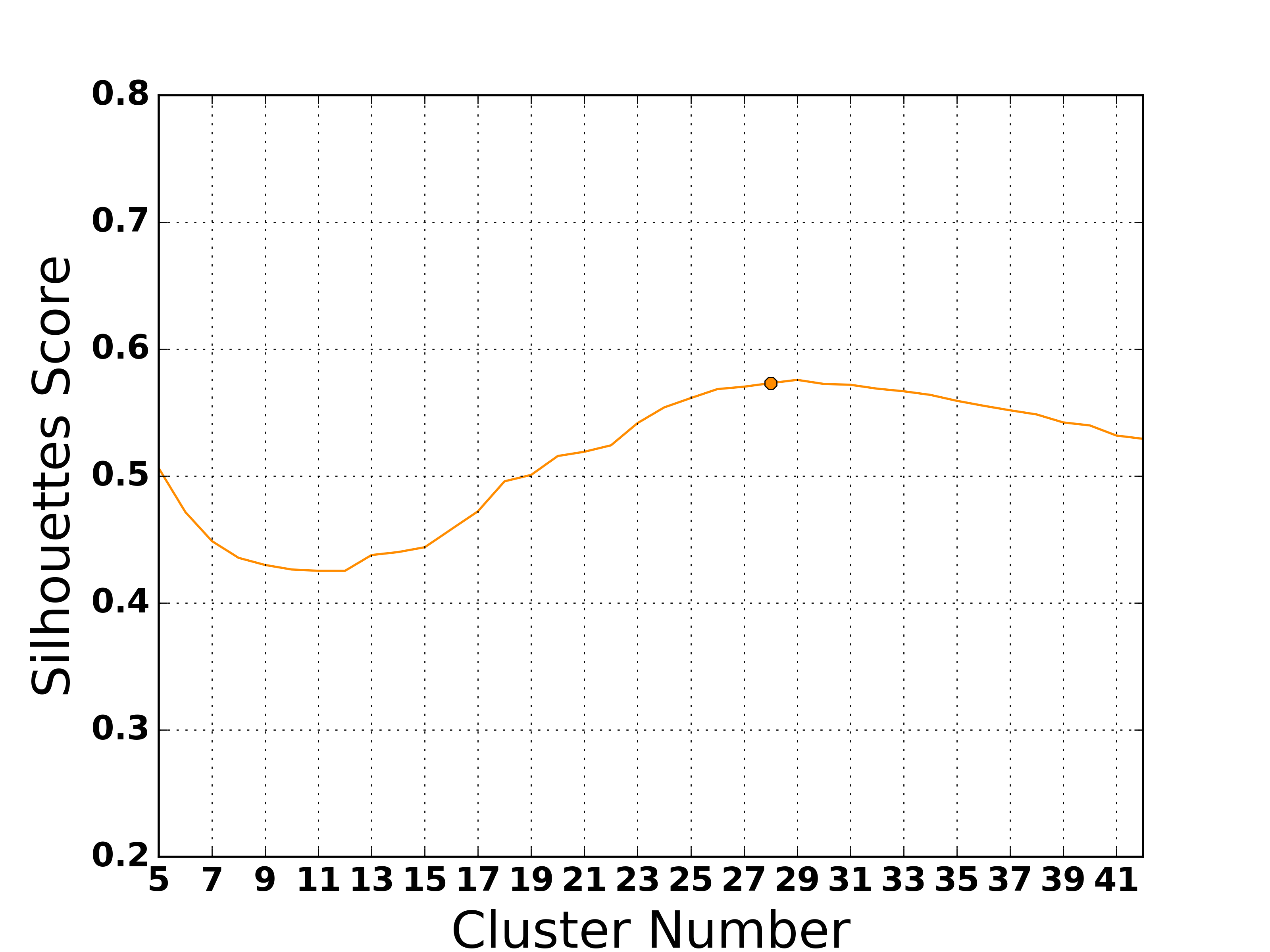}
  \caption{Silhouette coefficients for different numbers of clusters.}
  \Description{}
  \vspace{-0.3cm}
\end{figure}

From figure 8 we observed that when the number of clusters is 28, the clustering quality is the highest, so we select 28 as the number of clusters of long-term user behavior in the industrial data set.

\begin{figure}[h]
  \centering
  \includegraphics[scale=0.3]{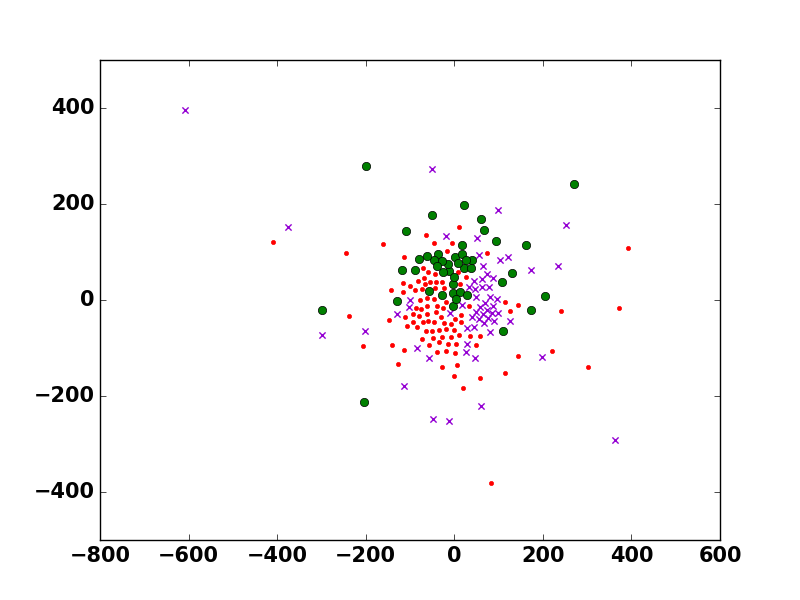}
  \caption{Three clusters visualization.}
  \Description{}
  \vspace{-0.3cm}
\end{figure}

\textbf{Clustering Visualization}:In Figure 9, in order to observe the clustering effect in GRS, we show the dimensionality reduction of three cluster centers, each of which represents a user's interest. We found that each interest contains a certain amount of behavior, and the differences between interests are more obvious.

\begin{figure}[h]
 \vspace{-0.1cm}
  \centering
  \includegraphics[scale=0.3]{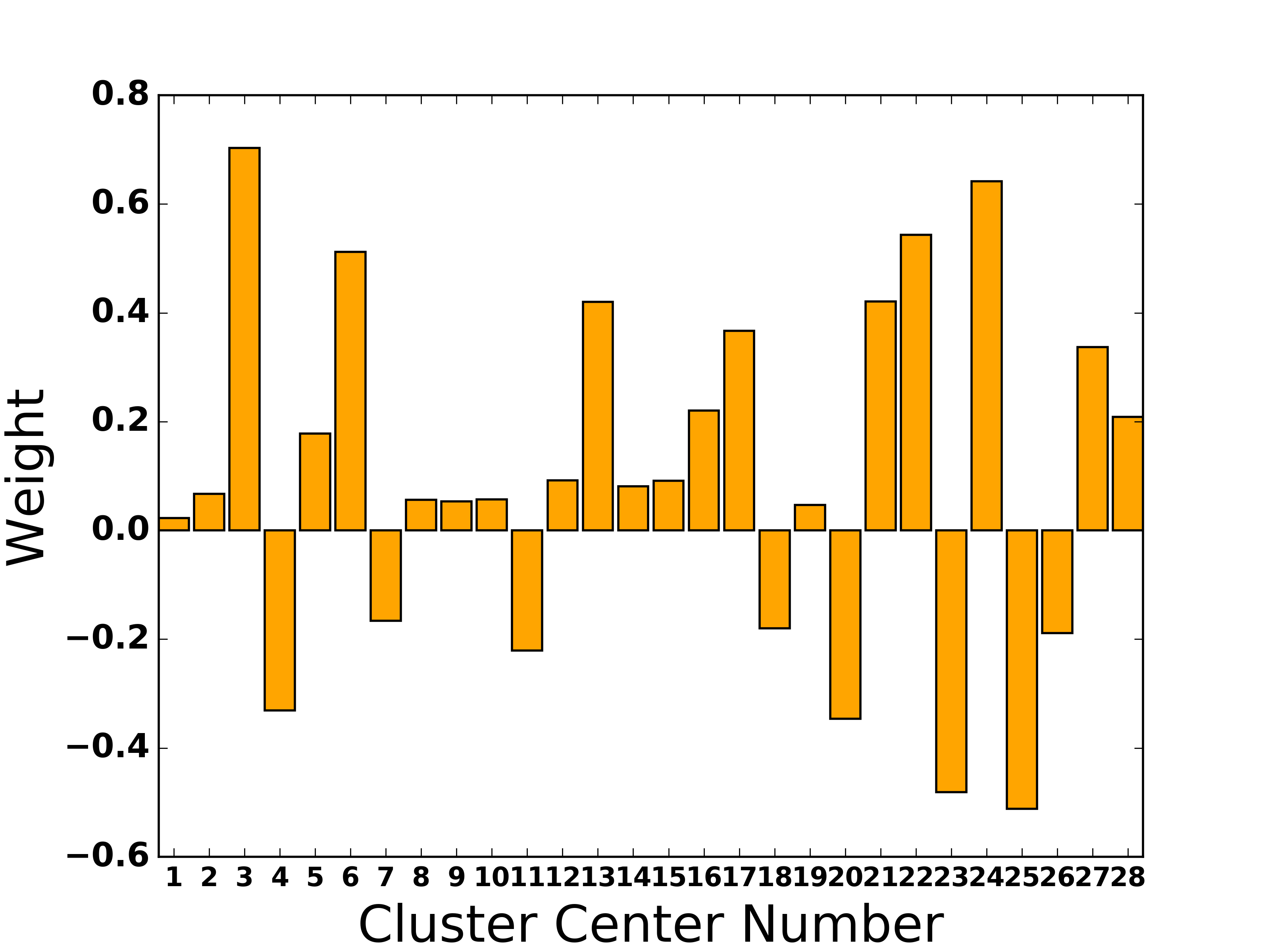}
  \caption{Attention weights of each interest cluster.}
  \Description{}
  \vspace{-0.3cm}
\end{figure}

\textbf{Multiple Interest Weights}: The importance of the association between target items and user interests is not the same. GLSM dynamically activates multiple interests through the interest activation unit. It can be seen from Figure 10 that during long-term interest matching, the weights of different interests have obvious personalized distributions.This matches people's common sense perception. For example, when the target item is a T-shirt, the relevance of clothing interests should be much higher than that of food interests.

\subsubsection*{\textbf{B. Short-term multi-intent recognition}}
\ 

The short-term behavior represents the user's recent interest distribution. Meanwhile, users’ short-term intentions usually include multiple scenarios (such as home, company, morning, noon), which consist of users’ ongoing behaviors. Our proposed short-term multi-intent is only able to capture the multi-scene interests of users. As can be seen from Figure 11, GLSM decreased exposure to soy milk (usually at breakfast) at lunchtime and increased exposure to braised chicken rice (usually at lunch) compared to the base. It can be seen from this data that GLSM can better fit the actual interests and preferences of users in different dining scenarios.

\begin{figure}[h]
  \centering
  \includegraphics[scale=0.3]{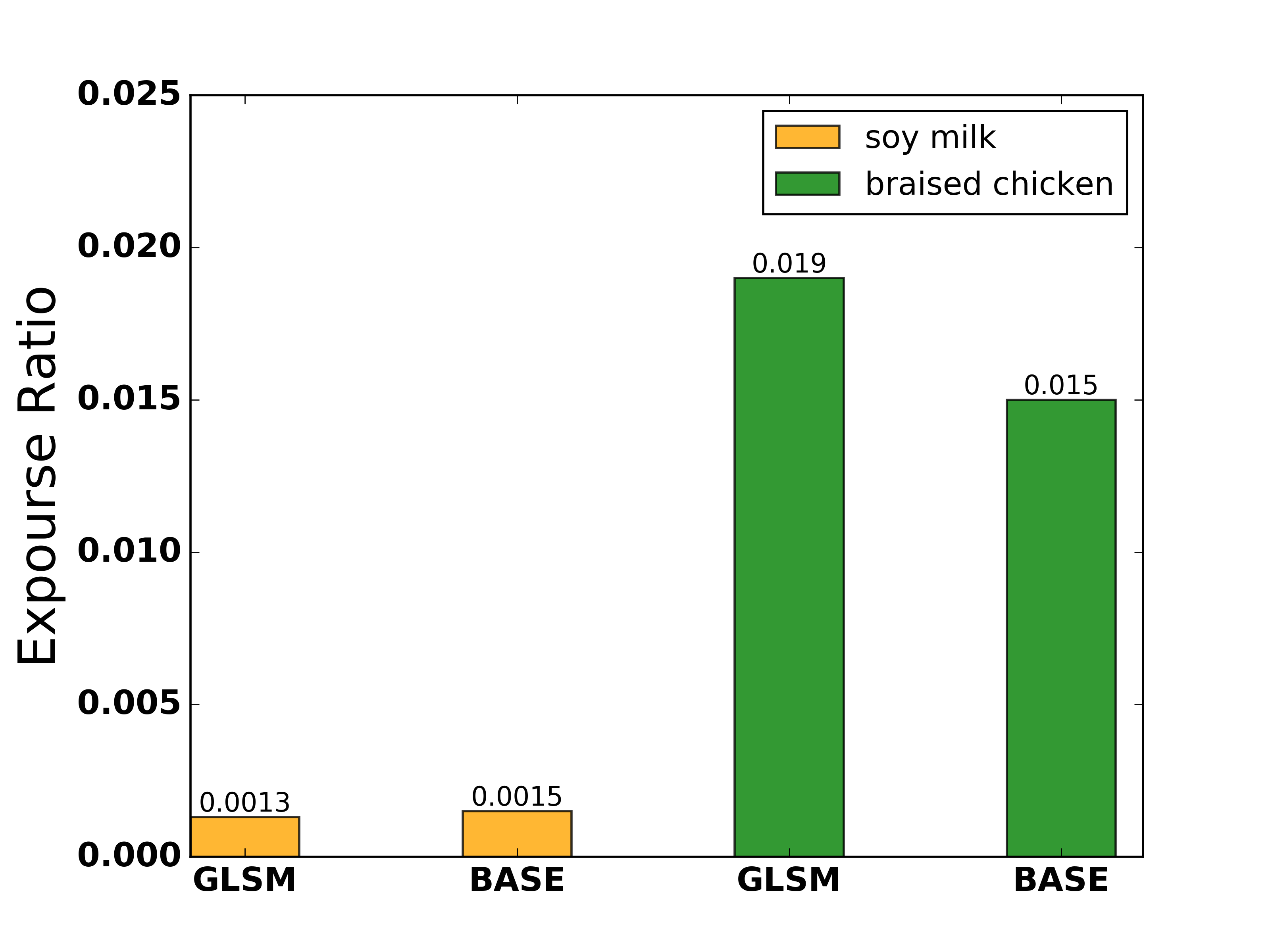}
  \caption{The exposure ratio of different types of food in the meal segment. Soy milk is usually served as a breakfast food, and braised chicken is usually served as a lunch food.}
  \Description{}
  \vspace{-0.3cm}
\end{figure}

\subsection*{RQ3: Ablation and Hyperparameter Studies}

\subsubsection*{\textbf{A.Numbers of topk center nodes in long-term module}}
For the long-term interest module of the GLSM model, we have analyzed the clustering quality for different numbers of clusters in the previous section. Here, we analyze the impact of choosing different topk center nodes on model performance. It can be seen from Figure 12 that with the increase of the number of TopK center nodes, the model effect shows a trend of first rising and then decreasing. Therefore, we set topk = 15 which means to select 15 center nodes from the center node set.

\begin{figure}[h]
  \centering
  \includegraphics[scale=0.3]{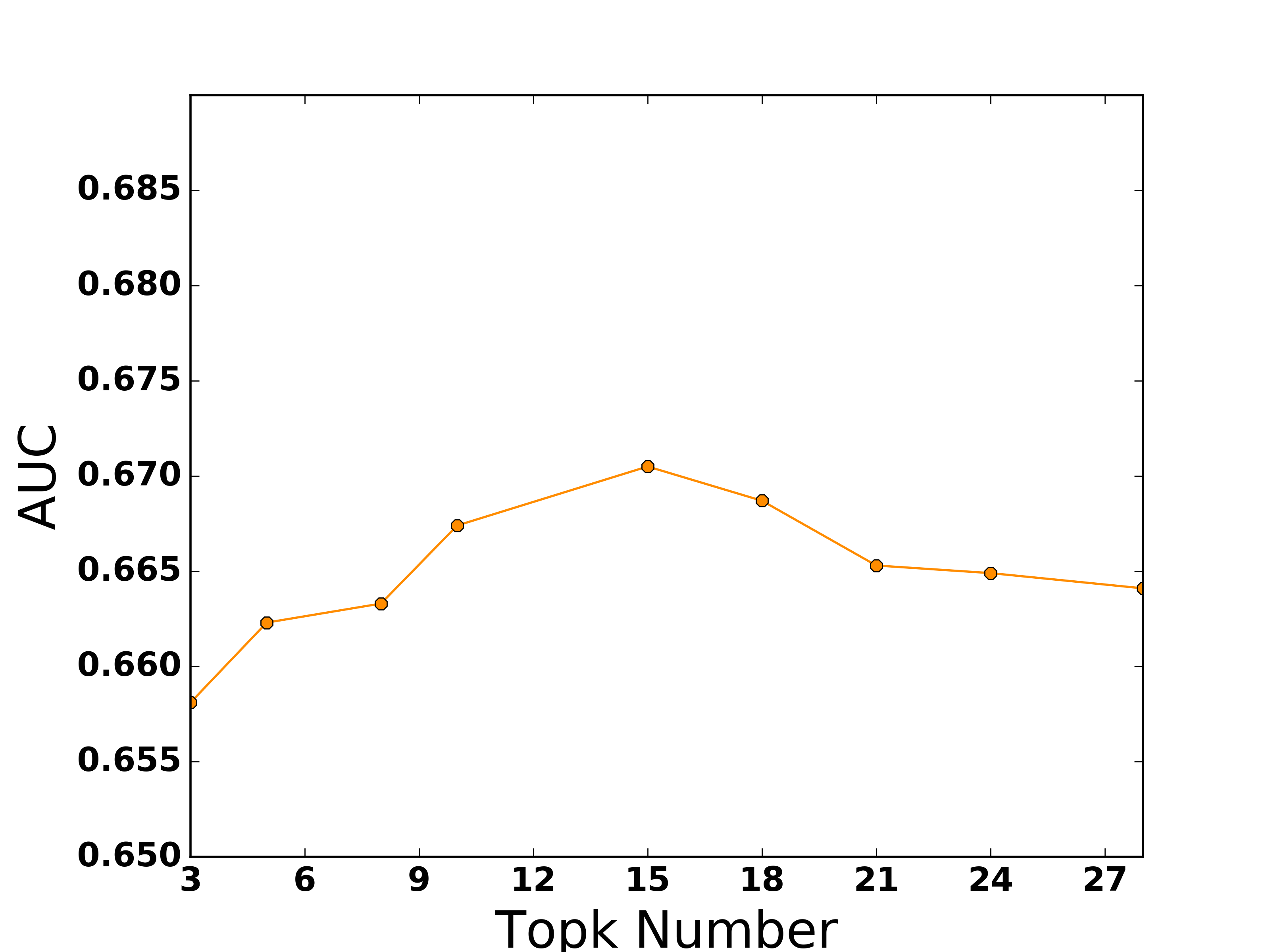}
  \caption{The effect of different TopK center node on model performance.}
  \Description{}
  \vspace{-0.3cm}
\end{figure}

\subsubsection*{\textbf{B.Long-term and short-term interests fusion}}

In GLSM, we propose User Personalized Gate Network to fuse users' long-term and short-term interests. And we also compare it with several common fusion methods. Table 3 shows the long-term and short-term fusion method we designed improves AUC by over 0.002 compared with add, weight, multiply and concat.

\begin{table}[h]
  \caption{Comparisons of different fusion methods}
  \label{tab:freq}
  \begin{tabular}{ccl}
    \toprule
    fusion&AUC&comments\\
    \midrule
    add & 0.6622 & sum each dimension\\
    weight & 0.6661 & weight sum with scalar\\
    multiply & 0.6664 & multiply each dimension\\
    concat & 0.6685 & concatenate \\
    gate fusion & 0.6705 & our method\\
  \bottomrule
\end{tabular}
\end{table}

In conclusion, we have conducted extensive experiments and multiple experimental comparisons, our proposed GLSM model shows significant results both offline and online.

\section{CONCLUSIONS}
In this paper, we focus on modeling users' long-term and short-term behavior. For long-term behaviors, we propose to build a graph retrieval structure to extract user interests and retrieve relevant long-term behaviors through center nodes. Compared with the SOTA baseline, we can extract various interests of users more personalized and reduce the information loss of long-term behavior. At the same time, graph retrieval can run in parallel with other processes to meet the online performance. For short-term behaviors, we split behaviors by scene to reduce confounding effects between scenes. In a single scene, GRU is used to extract the evolution of user interests, and in multiple scenes, user interests that match the target item are selected by an attention mechanism. For long Short-term fusion, the fusion network infers the influence of long-term and short-term behaviors individually based on user characteristics. Finally, we deployed the GLSM model on our platform. The GLSM model has brought significant business improvement and served mainstream traffic.



\bibliographystyle{ACM-Reference-Format}
\balance
\bibliography{ref}




\end{document}